\documentclass[twocolumn]{aastex63}
\usepackage{multirow}
\usepackage[T1]{fontenc}
\usepackage{mhchem}
\usepackage{url}

\usepackage{graphics,graphicx}

\newcommand{\Htwo}{\(\textrm{H}_{2}\)}
\newcommand{\CO}{\(^{12}\textrm{CO}\)}
\newcommand{\thCO}{\(^{13}\textrm{CO}\)}
\newcommand{\CetO}{\(\textrm{C}^{18}\textrm{O}\)}

\newcommand{\Msun}{\(\textrm{M}_{\odot}\)}

\newcommand{\CsvO}{\(\textrm{C}^{17}\textrm{O}\)}

\newcommand{\thCetO}{\(^{13}\textrm{C}^{18}\textrm{O}\)}

\definecolor{LightYellow}{rgb}{1,.98,.804}

\submitjournal{ApJ}

\begin{document}
\title{High Mass Inner Regions Found in Five Outbursting Sources}

\shorttitle{NOEMA Survey of Outbursting YSOs}


\author[0000-0002-0150-0125]{Jenny K. Calahan}
\affiliation{University of Michigan, 323 West Hall, 1085 South University Avenue, Ann Arbor, MI 48109, USA }
\affiliation{Center for Astrophysics | Harvard \& Smithsonian, 60 Garden St., Cambridge, MA 02138, USA }

\author[0000-0003-4179-6394]{Edwin A. Bergin}
\affiliation{University of Michigan, 323 West Hall, 1085 South University Avenue, Ann Arbor, MI 48109, USA }

\author[0000-0002-2555-9869]{Merel van 't Hoff}
\affiliation{University of Michigan, 323 West Hall, 1085 South University Avenue, Ann Arbor, MI 48109, USA }

\author[0000-0003-4001-3589]{Ke Zhang}
\affiliation{Department of Astronomy, University of Wisconsin-Madison, 475 N. Charter St., Madison., WI 53706}

\author[0000-0002-3950-5386]{Nuria Calvet}
\affiliation{University of Michigan, 323 West Hall, 1085 South University Avenue, Ann Arbor, MI 48109, USA }

\author[0000-0003-1430-8519]{Lee Hartmann}
\affiliation{University of Michigan, 323 West Hall, 1085 South University Avenue, Ann Arbor, MI 48109, USA }

\begin{abstract}
Young stellar objects are thought to commonly undergo sudden accretion events that result in a rise in bolometric luminosity. These outbursts likely coincide with the onset of planet formation, and could impact the formation of planets. The reason behind this dramatic enhancement of accretion is an active area of research, and the mass of the system is a critical parameter. Using Northern Extended Millimeter Array, we survey five outbursting sources (three FU Ori, one EX Or, one `peculiar' source) with the primary goal of determining the system's mass using an optically thin line of CO. We estimate the mass of a central region for each object that using both continuum emission and \CsvO{} J=2-1. The \CsvO{} emission likely includes both disk and inner envelope material, thus acts as an upper limit on the disk mass, ranging from 0.33-3.4 M$_{\odot}$ for our sources. These derived masses suggest that the inner $\sim$1000 au contains enough mass along the line of sight for these sources to be gravitationally unstable.  

\end{abstract}


\section{Introduction} \label{sec:intro}

FU Orionis-type objects (FU Ori) and EX Ori-type objects (EXOr) have been theorized to be a common evolutionary phase between the Class I and Class II stages for low-mass pre-main sequence stars \citep{Hartmann96,Quanz07,Vorobyov15}. An FU Ori object is classified as such after undergoing a sudden and extreme (several magnitudes) increase in brightness in optical and near-infrared wavelengths \citep{Herbig77}. This outburst has been attributed to short-lived and massive accretion events, predicted to accrete up to 10$^{-4}$ -- 10$^{-5}$~\Msun{}/yr \citep{Hartmann96,Audard14,Fischer23} onto the star. While ExOrs also undergo sudden increases in observed magnitude, they are less extreme and outbursts happen more frequently than FU Oris. There are a number of mechanisms that could cause this sudden and fast accretion, including thermal or gravitational instabilities \citep{Bell94,Armitage01,Boley06,Zhu10}, perturbations from a close-by massive companion \citep{Bonnell92}, or sporadic infall from the circumstellar envelope \citep{Vorobyov06, Vorobyov13}. Robust mass constraints on the disk and inner envelope are essential in determining what mechanism(s) is/are responsible for FU Ori outbursts. Gravitational instabilities are associated with triggering a magnetorotational instability (MRI) which would cause the sudden burst of accretion, and this would require the disk to be relatively massive (M$_{disk}$/M$_{*} \gtrsim$ 0.1) \citep{Hartmann96,Liu16,Cieza18} and thermal instabilities require regions of the disk that have high enough opacity to trap heat \citep{Zhu09}. Regardless of the mechanism, FU Ori systems need to be massive enough to sustain the high accretion rate for many years. 

CO is commonly used as a mass tracer for molecular clouds, protostellar envelopes, and protoplanetary disks, as it is a highly abundant molecule, is chemically stable, and easily detectable. The measured ratio of CO/\Htwo{} = $10^{-4}$ in the ISM \citep[i.e.][]{Frerking82,Kramer99,Parvathi12}, and is consistent with the ratio found in young Class 0/I disks \citep{van'tHoff20,Zhang20}. 
FU Ori objects present a unique laboratory for the purpose of mass determination of young stellar systems due to their recent outbursts. The outburst heats up the surrounding disk and envelope, pushing the snowlines beyond their regular radial extent \citep{Cieza16,vantHoff2018a}, exposing more of the disk's mass to be probed with CO. 
The radial location of the CO sublimation front (temperature $\approx$~20~K) during an FU Ori outburst can almost quadruple or even increase tenfold (at the intermediate stage of luminosity rise, and the outburst maximum, respectively) \citep{Molyarova18}. 

\CO{} emission is found to be optically thick towards typical molecular clouds and protoplanetary systems, thus can only provide an lower limit to the mass. Given this, we use lesser abundant isotopologues of \CO{} to probe denser regions. \thCO{} and \CetO{} have often been used to probe higher density regions. \CetO{} emission (both the $J$= 1-0, and 2-1 transitions) is generally assumed to be optically thin, and has been used as the mass tracer for FU Ori objects and other young stellar objects. Recent work shows, however, that \CetO{} emission may not be completely optically thin. Available observations of FU Ori objects find that \CetO{} $J$=1-0 can have measured $\tau > 1$ \citep{Feher17}. Towards those same disks, emission from the brighter and more abundantly populated \CetO{} $J$=2-1 line is expected to have an even higher $\tau$. Recent observations with Northern Extended Millimeter Array (NOEMA) towards three non-FU Ori Class I objects showed that \CetO{} and \CsvO{} $J$=2-1 were both optically thick in the disk region, and \thCetO{} was required to reach optically thin emission levels \citep{Zhang20}. 

To determine the total mass of the inner regions of these outbursting source, an optically thin tracer (\CsvO{} or \thCetO{}) is used, along with the measured temperature (derived from \CetO{}) and an assumed abundance of the optically thin tracer to \Htwo{}. 
We present NOEMA observations of five outbursting sources with the primary science targets of \ce{^{13}CO}, \ce{C^{18}O}, \ce{C^{17}O}, and \ce{^{13}C^{18}O}. We will use the optically thin tracer to estimate the mass for each source. We focus on the central region, as defined by the peak of the continuum, covering areas with equivalent radii of hundreds of au. 

\section{Methods}\label{sec:method}
\subsection{Source Selection}
For our sample, we isolate sources that have been published and observed with NOEMA or ALMA in \CetO{} $J$=1-0 or 2-1, and are located within the observational range of NOEMA \citep{Feher17,Principe18,Abraham18}. There are nine total sources that fit this criteria; we chose the brightest five sources, each with \CetO{} emission peaking directly on source. Within the northern hemisphere, these objects have the highest likelihood of a \CsvO{} and \thCetO{} detection. Source properties can be found in Table 1. Our sources range from classic FU Ori objects (V1057 Cyg, V1735 Cyg, V1515 Cyg), to a peculiar source with unique spectral features but otherwise FU Ori characteristics (V1647 Ori), to an EXor/UXor type star with a higher period of luminosity variability (V2492 Cyg). 

V1057 Cyg and V1515 Cyg, along with the namesake source FU Ori were the first three discovered and defined the class of FUOr-type objects \citep{Herbig77}. V1735 Cyg was discovered later, and exhibited the characteristic sharp rise in luminosity followed by a slow decay with short-period modulations \citep{Szabo22}. V1647 Ori was at first categorized as an FU Ori type object due to its extreme rise in luminosity, however has exhibited multiple significant outburst events since the first one captured, thus it has been classified as either borderline FU Ori/EX Or, or `peculiar' source \citep{Audard14, Connelley18, Fischer23}. V2492 Cyg was seen to go into an outburst stage in 2010, and since then its variability in brightness has been attributed to a combination of episodic changes in extinction and accretion \citep{Hillenbrand13, Giannini18, Ibryamov21}.

\begin{deluxetable*}{ccccccc}
\label{tab:SourceProperties}
\tablecolumns{5}
\tablewidth{0pt}
\tablecaption{Source Properties}
\tablehead{Name & Classification & R.A. & Dec & Distance [pc] & Outburst Date & L$_{\rm bol}$ Peak [L$_{\odot}$]}
\startdata
V1057 Cyg	&	FU Ori	& 20h 58m 53.733s	& +44d 15\arcmin~ 28.389\arcsec & 907$^{+19}_{-20}$	&	1970	&	250-800	\\
V1735 Cyg	&	FU Ori	& 21h 47m 20.663s & +47d 32\arcmin~ 3.857\arcsec &	691$^{+35}_{-38}$	&	1957-1965	&	235	\\
V1647 Ori	&	Peculiar & 5h 46m 13.137s & -0d 6\arcmin~ 4.885\arcsec	&	413$^{+20}_{-22}$	&	1966, 2003, 2008	&	34-44	\\
V1515 Cyg	&	FU Ori	& 20h 23m 48.016s & +42d 12\arcmin~ 25.781\arcsec&	902$^{+12}_{-13}$ 	&	1950	&	200	\\
V2492 Cyg	&	UX Ori/EX Ori	& 20h 51m 26.237s	& +44d 05\arcmin~23.869\arcsec & 804$^{+24}_{-26}$	&	2009-2010	&	43	\\
\enddata
\tablecomments{Distances taken from parallaxes in Gaia Release 3 \citep{Gaia3}, however it is worth noting that outburst soruces are particularly difficult to determine distances to. Outburst date and luminosity taken from review \cite{Audard14} and references therein. Stated ranges indicate the timespan where the onset of the outburst could have started.}
\end{deluxetable*}

\subsection{NOEMA Data Reduction}

These results are derived from NOEMA observations that took place July - October 2021 as a part of project S21AL. Each source was observed for $\approx$3.4 hrs with beam sizes $\approx$ 1.$^{\prime\prime}$58 $\times$ 1.$^{\prime\prime}$50 (V1735 Cyg, V1057 Cyg) 2.$^{\prime\prime}$45 x 1.$^{\prime\prime}$69 (V1647 Ori) 1.$^{\prime\prime}$04 $\times$ 0.$^{\prime\prime}$92 (V2492 Cyg, V1515 Cyg). We utilized Band 3, specifically in the range of 202-211 GHz (lower sideband) and 218-226 GHz (upper sideband). This setup covers our key science lines, simultaneously covering the \thCO{}, \CetO{}, \CsvO{}, and \thCetO{} J=2-1 transitions. The initial round of data quality assurance and reduction was done with a staff member at IRAM (Institut de Radioastronomie Millimetrique). The raw data was self-calibrated and CLEAN-ed using the GILDAS package \textsc{mapping}\footnote{https://www.iram.fr/IRAMFR/GILDAS}. The first step was to find the systematic velocity and correct the UV-tables to the accurate values, which aids in line detection. Next we isolate the continuum by zero-ing out every line and other feature such as dead pixels or end-of-band wiggles. With the channels containing only continuum emission left over, we average over every $\sim$200 channels before self-calibration and imaging. 
That solution is then imposed on the original spectra that contains the emission lines. The self-calibrated data is then continuum-subtracted and imaged. For the goal of determining the mass of each source, we utilized the high-velocity resolution (0.5 km/s or 62.5 kHz) spectra of \ce{^{13}CO}, \ce{C^{18}O}, \ce{C^{17}O} and \ce{^{13}C^{18}O}. V1735 Cyg likely contains significant extended emission in the three brightest CO isotopologues, as its solution contains artifacts that proved to be less strong if the smallest baselines were excluded. However, to maintain consistency in the analysis of all sources, the full NOEMA configuration including all baseline was used for the final results. 


\section{Results}\label{sec:results}

\begin{figure*}
    \centering
    \includegraphics[scale=.45]{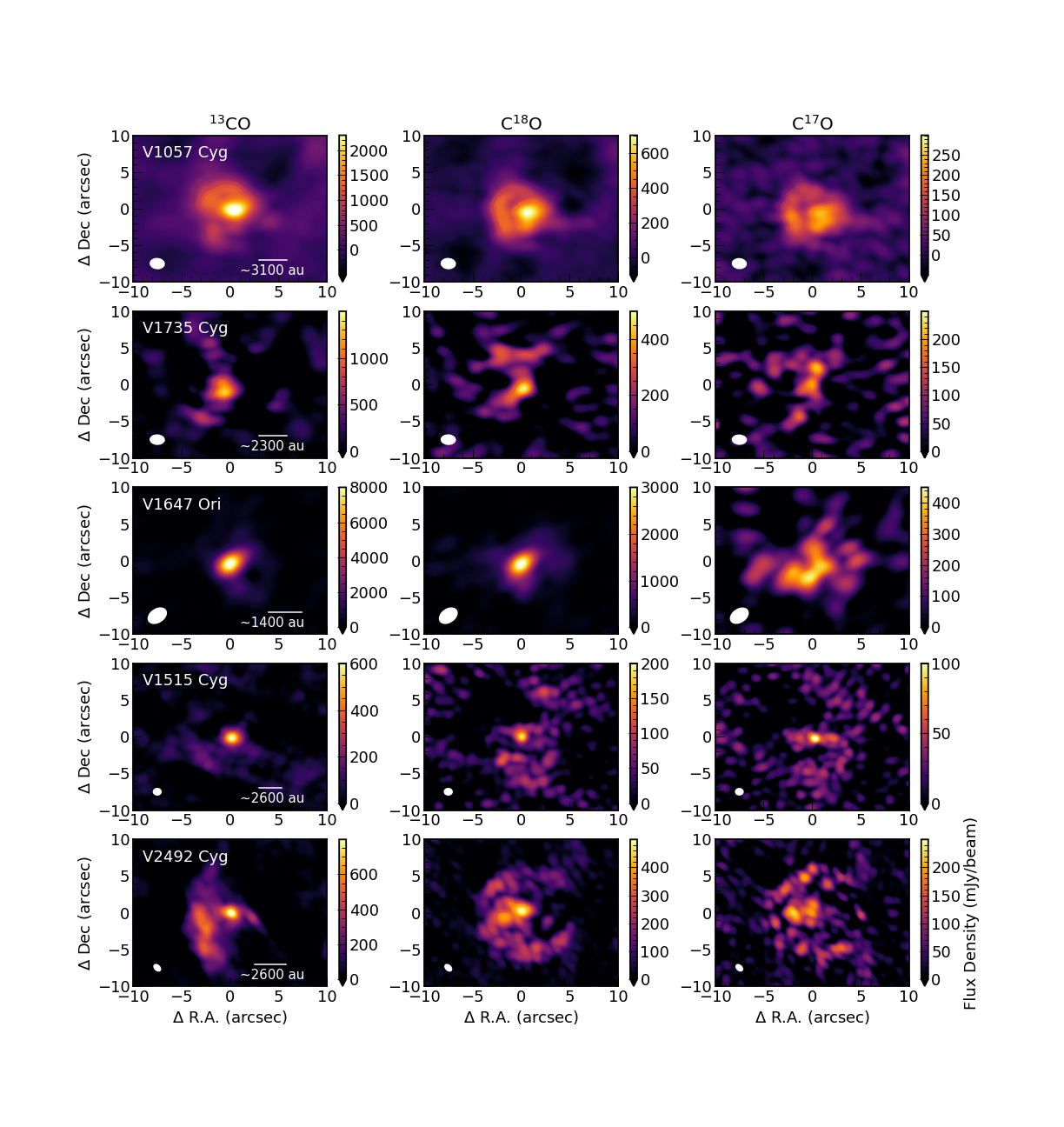}
    \vspace{-.75in}
    \caption{Moment zero maps of \thCO{}, \CetO{}, \CsvO{} J=2-1 for each source observed in this survey. }
    \label{fig:all_moments}
\end{figure*}

In our NOEMA sample we have observed \ce{^{13}CO}, \ce{C^{18}O}, and \ce{C^{17}O} towards all five sources, and the emission is partially to fully resolved. Towards one of them, V1057 Cyg, we also have an unresolved detection of \ce{^{13}C^{18}O}. In the moment zero maps, all sources have \thCO{} and \CetO{} peaked on source, coincident with the continuum. \CsvO{} peaks on source for V1057 Cyg and V1515 Cyg, however the other three have \CsvO{} surrounding or nearby the central star. It is likely that the continuum may be blocking some central \CsvO{} emission for these three sources. In all sources, there is extended gas emission as compared to the continuum (with V1515 Cyg being being the most compact, see Figure \ref{fig:all_moments}).  

We seek to determine the mass of the central region from both the continuum observations and an optically thin gas line. For the continuum mass derivation, we use the following equation:

\begin{equation}
    M_{cont} = \frac{S_{\nu}d^{2}}{\kappa_{\nu}B_{\nu}(T)}.
\end{equation}

\noindent Here, $S_{\nu}$ is the flux density, d is the distance, $B_{\nu}({T})$ is the Plank function for a given dust temperature T, and $\kappa_{\nu}$ is the dust opacity coefficient which we assume to be 2.2 cm$^{2}$g$^{-1}$ \citep[$\kappa_{\nu} = 10(\nu/1000 ~\rm{GHz})$ ;][]{Beckwidth90, Cieza18}. For the dust temperature, we uniformly use T=25~K for both the dust and gas mass determinations, as that is approximately the brightness temperature for the continuum of each object. We use a gas-to-dust mass ratio of 100, and thus find inner-region masses of 0.39-0.85 M$_{\odot}$ (see Table \ref{tab:SourceResults}).

Our ideal gas-mass tracer will be an optically thin transition of CO, and so we first calculate the optical depths of each of these isotopologues. The brightness temperature of a molecule is defined by:

\begin{equation}
    T_{B} = T_{ex} (1-e^{-\tau})
\end{equation}

\noindent where $T_{ex}$ is the excitation temperature and $\tau$ is the optical depth.  Taking the ratio between the brightness temperatures of two isotopologues will provide a relationship between the optical depths of the two lines:

\begin{equation}
    \frac{T_{B}(^{13}CO)}{T_{B}(C^{18}O)} = \frac{T_{ex}(^{13}CO) (1-e^{-\tau_{13}})}{T_{ex}(C^{18}O) (1-e^{-\tau_{18}})}.
\end{equation}

\noindent The excitation temperatures of each isotopologue of the same rotational transition can be assumed to be equal. And if you can assume that the \ce{^{13}CO} is optically thick, you can simplify this relation to

\begin{equation}
    \frac{T_{B}(^{13}CO)}{T_{B}(C^{18}O)} \approx \frac{1}{1-e^{-\tau_{18}}}.
\end{equation}

\noindent One can further state that $\tau_{18} $ = $\tau_{13}/X$ where X is the abundance ratio between the two isotopologues. Thus, the ratio of the optical depths is equal to the abundance ratio. In our case we use ratio values from \citet{Qi11} which were found from models that reproduce multiple observations of CO isotopologues in a protoplanetary disk, and are consistent with the ISM \citep{Wilson99}: \ce{^{12}C}/\ce{^{13}C}=67, \ce{^{16}O}/\ce{^{18}O}=444, and \ce{^{18}O}/\ce{^{17}O}=3.8. If \(R=\frac{T_{B}(^{13}CO)}{T_{B}(C^{18}O)}\) then you can express the optical depth of each isotopologue line as follows:

\begin{equation}
    \tau_{13} = -\frac{444}{67} \ln(1-\frac{1}{R}) 
\end{equation}

\begin{equation}
    \tau_{18} = -3.8 \ln(1-\frac{1}{R})
\end{equation}

\begin{equation}
     \tau_{17} = \tau_{18}/3.8   
\end{equation}

Using the observed flux ($S_{\nu} \propto T_{B}$) of each isotopologue, we calculated the optical depths for each observed line. We started with moment zero maps derived from our cleaned spectral cubes using bettermoments \citep{Teague18}. 
All five sources show optically thin emission in the \ce{C^{17}O} transition, while \ce{C^{18}O} is marginally optically thick ($\tau \approx$ 1 in most sources), and all \ce{^{13}CO} is optically thick. 

\begin{deluxetable}{cccc}
\label{tab:SourceResults}
\tablecolumns{7}
\tablewidth{0pt}
\tabletypesize{\small}
\tablecaption{Observational Properties}
\tablehead{Source	& $\theta_{maj}$ & $\theta_{min}$ & P.A.
	 }
 \startdata
 V1057 Cyg & 1.577 & 1.499 & 58.9\\
 V1735 Cyg & 1.575 & 1.412 & 82.7\\
 V1647 Ori & 2.458 & 1.689 & -33.6\\
 V1515 Cyg & 1.041  & 0.920 & 1.80\\
 V2492 Cyg & 1.115 & 0.715 & 26.9\\
 \enddata
\end{deluxetable}

\begin{deluxetable*}{ccccccc}
\label{tab:SourceResults}
\tablecolumns{8}
\tablewidth{0pt}
\tabletypesize{\small}
\tablecaption{Derived Column Densities and Mass}
\tablehead{Name	&	log$_{10}$(N$_{C17O \rm avg}$) 	&	log$_{10}$(N$_{H2 \rm avg}$) 	&	Emitting Radius	& Mass$_{dust} \times 100$	&	Mass$_{C17O}$ & Toomre Q 	\\
	&	log$_{10}$([$\frac{\rm mol}{\rm cm^2}$])	&	log$_{10}$([$\frac{\rm mol}{\rm cm^2}$])	&	[au]	&	[M$_{\odot}$]	& }
\startdata
V1057 Cyg	&	17.3	&	24.5	&	922	&	0.80	&	3.4	& 0.06\\
V1735 Cyg	&	16.3	&	23.5	&	769	&	0.39	&	0.25	& 0.7\\
V1647 Ori	&	17.2	&	24.4	&	379	&	0.64	&	0.42 & 0.3	\\
V1515 Cyg	&	16.7	&	24.0	&	726	&	0.47	&	0.58	& 0.3\\
V2492 Cyg	&	17.8	&	25.1	&	267	&	0.85    &	1.9 & 0.07\\
\enddata
\end{deluxetable*}

The mass of each source is therefore calculated using \ce{C^{17}O} emission. In an environment that is optically thin and in LTE, the column density of the upper energy state of a certain molecular transition is defined as:  

\begin{equation}
    N_{u}^{\rm thin} = \frac{4 \pi S_{\nu} \Delta \nu}{A_{ul} \Omega \rm{hc}},
\end{equation}

\noindent where $S_{\nu} \Delta \nu$ is the integrated flux density. The frequency range, $\Delta \nu$, is defined by the width of the \ce{C^{17}O} emission profile in frequency space, and we use the same frequency range for each source's mass estimate. $A_{ul}$ is the Einstein-A coefficient that represents the rate with which an energy level is depopulated through spontaneous emission, and this is a constant for the \ce{C^{17}O} J=2-1 transition and equal to $10^{-6.2}$ $\rm s^{-1}$ \citep{Muller05}. $\Omega$ is the solid angle of the emitting area.

After the column density of the J=2 level is calculated, an LTE assumption can easily relate it to the total \CsvO{} column density, $N_T$:

\begin{equation}
    N_{T} = \frac{N_{u}}{g_{u}} Q(T_{rot})[e^{-E_{u}/kT}]^{-1},
\end{equation}

\noindent where $g_{u}$ is the statistical weight of the J=2 level, $Q(T_{rot})$ is the partition function, and $E_{u}$ is the energy of the J=2 level. All of these constants are found in the JPL line catalogue \citep{Pickett98}. 

\begin{figure*}
    \centering
    \includegraphics[width=\textwidth]{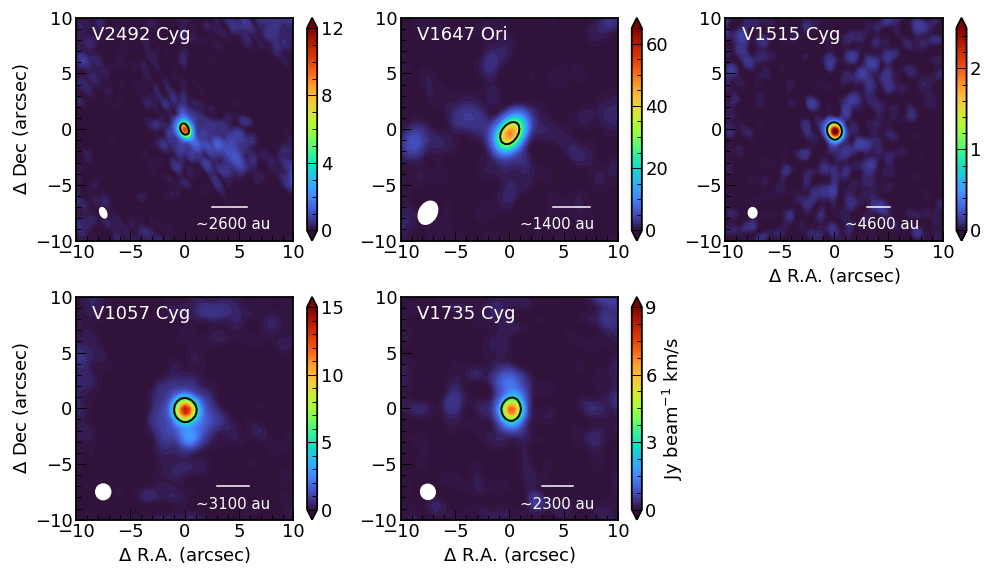}
    \caption{\textit{Continuum maps corresponding to $\sim$221 GHz, extracted from the upper sideband of our NOEMA observations. The bulk of the continuum emission resides within the black contours, which is used to calculate a continuum mass. All contours correspond to at least where 15$\sigma$ emission is found, and the corresponding areas for each source are similar spatial scales. These areas are used to calculate a gas mass from \CsvO{}, see Figure \ref{fig:disk_cloud}. } }
    \label{fig:disk_continuum}
\end{figure*}

\begin{figure*}
    \centering
    \includegraphics[width=\textwidth]{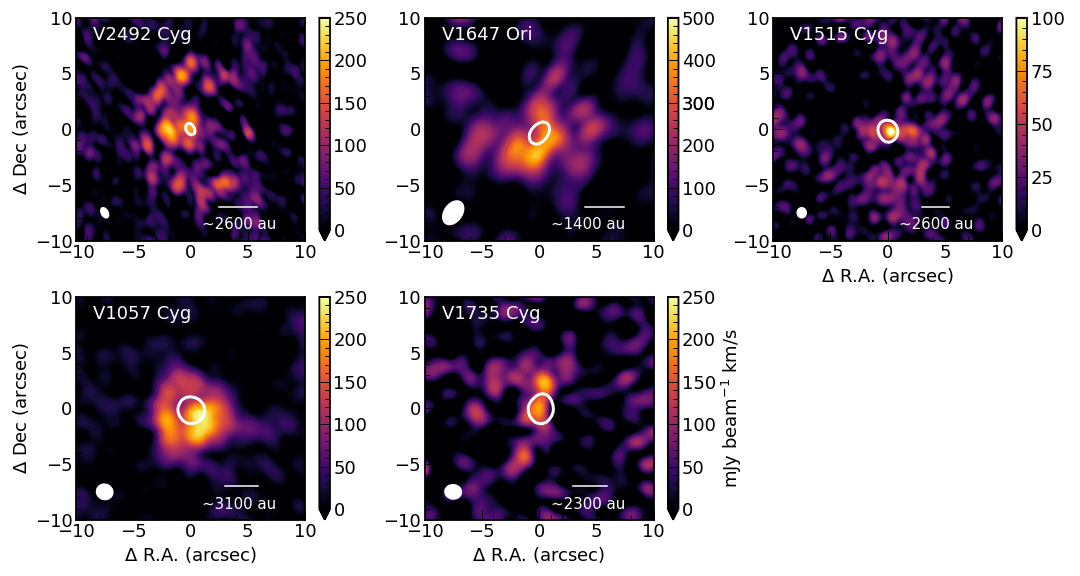}
    \caption{\textit{Moment zero maps of \ce{C^{17}O} with the white contour indicating the central area from which a mass is determined, as defined by the continuum flux. }}
    \label{fig:disk_cloud}
\end{figure*}

Once the total \ce{C^{17}O} column density is found, we can use the isotope ratios to back out the total CO column and then \ce{H2} column density. We use \ce{C^{16}O}/\ce{C^{18}O} = 1687 and CO/\ce{H2} = 10$^{-4}$ \citep{Wilson99,Frerking82,Bergin17}. Once the \ce{H2} column density is derived per pixel, we sum over the surface area of a set of given pixels to produce a total mass. Using the continuum data, we define a central area of the image that contains at least all of the 15$\sigma$ continuum emission (see Figure \ref{fig:disk_continuum}). Three sources (V1057 Cyg, V1515 Cyg, and V1647 Ori) only included 20$\sigma$ flux so that the central areas were comparable sizes between sources. The final calculated masses from \CsvO{} range between 0.25-3.4 M$_{\odot}$. These regions likely contains both disk and inner envelope emission in the corresponding \CsvO{} image (See Figure \ref{fig:disk_cloud}), in particular, emission from the envelope along our line of sight. All moment zero maps were used by implementing the \textsc{bettermoments} packages \citep{Teague18}.

\begin{figure}
    \centering
    \includegraphics[width=.45\textwidth]{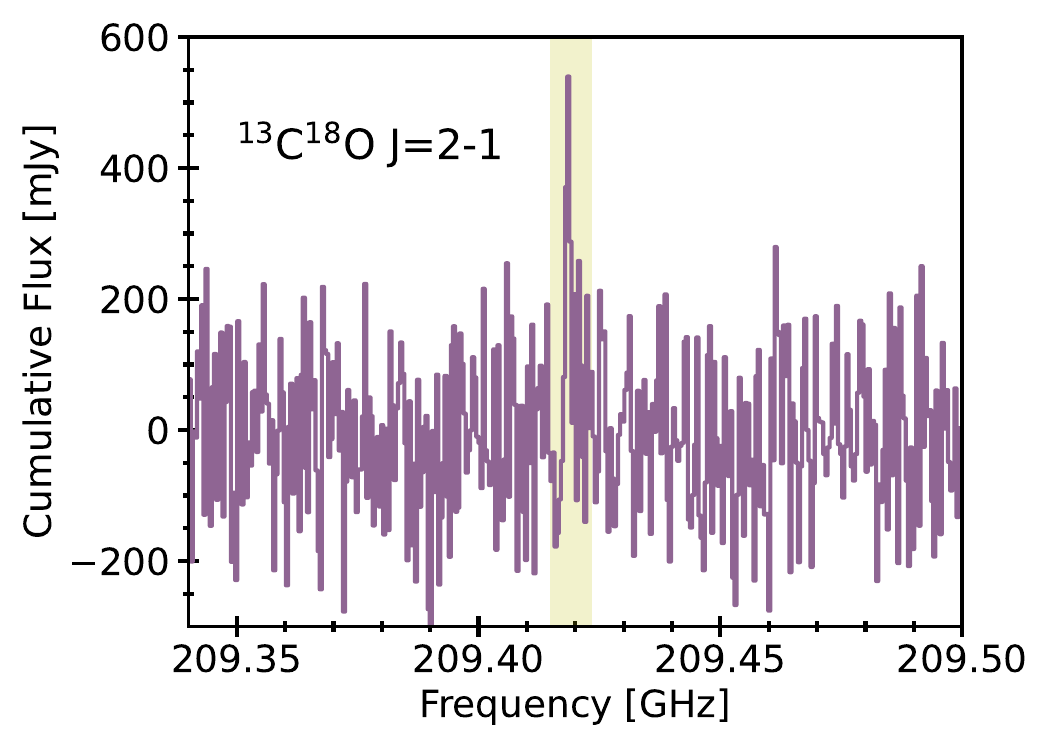}
    \caption{The $^{13}$C$^{18}$O J=2-1 detection towards V1057 Cyg, summed over the central few pixels with an equivalent area to the beam.}
    \label{fig:13c18o}
\end{figure}

The average \ce{C^{17}O} column densities, mass column densities, derived mass from \CsvO{}, continuum flux, the corresponding equvalent emitting radius, and the ratio between the final gas and dust masses are shown in Table \ref{tab:SourceResults}. These derived masses from the \CsvO{} are likely upper estimate on the mass of the disk, while also being lower-end estimates on the total mass within this emitting area as not all of our sources have their \CsvO{} peak overlap with the continuum peak.  

For the source V1057 Cyg, we also have a detection of $^{13}$C$^{18}$O which can be used as another mass tracer. The $^{13}$C$^{18}$O flux is not located in the central pixel, but is detected after summing over the area of the beam (see Fig. \ref{fig:13c18o}). We find an inner-region mass of 2.6 M$_{\odot}$ using $^{13}$C$^{18}$O. This used the sum of the flux over the same area used for the mass determination from both the continuum and C$^{17}$O, and a $^{12}$C$^{16}$O/$^{13}$C$^{18}$O = 29,748. Compared to the mass determined from C$^{17}$O, it is slightly lower however well within a factor of two, thus we interpret this to be consistent. Both optically thin tracers suggest V1057 Cyg to be the most massive source within our sample. 

\section{Discussion}\label{sec:diss}

Mass is a fundamental property when it comes to understanding the protoplanetary disk environment and FU Ori outburst events. The quickest and most often used technique to extract disk masses is using dust thermal continuum observations. Assuming optically thin dust emission and a gas-to-dust ratio of 100, a gas mass can be calculated from the dust observations. We use this technique to estimate a lower-end mass of the central region, due to the fact that this continuum emission is likely optically thick \citep[i.e.][]{Tobin20, Kospal21}. Within similar objects, the dust that is probed at $\sim$220 GHz has been shown to be optically thick, and corresponding \CsvO{} emission often shows a dip corresponding with the center of the continuum emission (see V1647 Ori and V1057 Cyg in Figure \ref{fig:disk_cloud}).  Using optically thin observations of C$^{17}$O, we also calculate a mass of the central region, and treat this as an upper-end estimate of the disk mass. In each of our sources, C$^{17}$O emission is optically thin, however it comes with some caveats. The emission likely probes the mass of the in-falling envelope and disk while the continuum probes large dust grains confined to the disk. Thus, the \CsvO{} emission likely probes more mass over a larger area, especially along the line of sight. However, the dust emission is expected to be optically thick and blocking C$^{17}$O emission from the backside of the disk/envelope. \CsvO{} will be an upper limit if a source has an envelope, however if there is no envelope the \CsvO{} mass could be closer to a lower limit given an optically thick dust disk. Three of our disks have total inner-region masses that have similar values when derived from the continuum (with gas/dust =100) and from an optically thin gas tracer (\CsvO{}). V1735 Cyg, V1647 Ori, and V1515 Cyg have masses that agree within a factor of less than two, suggesting that there may be little envelope contribution. V2492 Cyg, our EXOr source, has a gas mass that is just over 2 times more massive than what would be assumed from continuum observations alone. The final gas mass along the line of sight with a radial extent of  $\sim$1000~au around V1057 Cyg is 3.4 M$_{\odot}$, over four times more massive than what would be assumed from continuum alone. V2492 Cyg and V1057 Cyg likely have a significant contribution from the surrounding envelope, perhaps feeding the disk and playing a role in the mechanism that caused an episodic accretion event for these sources. 



Within our small sample, there appears to be no trend in mass versus outburst classification, peak luminosity, nor most recently observed brightness \citep[most up-to-date photometic observations for each source:][]{Kopatskaya13, Peneva09, Ninan13, Szabo22, Ibryamov21}. Our two most massive sources are an FU Ori (V1057 Cyg) and EX Ori object (V2492 Cyg) which exhibit markedly different peak luminosities and time since outburst. The fact that these two sources are the two most massive in our sample agrees with previous work that measured the mass of these two sources \citep{Feher17} but does not agree with a trend seen when observing the continuum probed with ALMA that EX Ori sources tend to be less massive than FU Ori sources \citep{Cieza16}. The most thorough samples of FU Ori masses come from continuum emission alone, with $\approx$18 determined masses \citep[][ and references therein]{Cieza18, Kospal21}. Compared to previously observed sources, our five sources reside alongside the most massive that have been seen with ALMA: V883 Ori, Haro 5a IRS, V900 Mon, and L1551 IRS 5 N. Our sample was biased to target the brightest outbursting sources in the Northern Hemisphere, thus it is not particularly surprising that we find higher masses than the average source in \citet{Cieza18,Kospal21}. Direct comparisons of the continuum mass are not straightforward between papers, as different emitting areas are used, temperatures, and distances. However, it is insightful to determine if our sample hold true with a trend seen thus far, that the majority of FUOr sources are gravitationally unstable.   

The Toomre Q parameter is often used to determine if a disk is gravitationally stable or not \citep{Toomre64}. A Q parameter greater than one suggests the disk is stable, while less than one is unstable. We use the following expression to calculate the Toomre Q parameter:

\begin{equation}
    Q \equiv \frac{c_s \Omega}{\pi G \Sigma}.
\end{equation}

The angular speed of the gas disk ($\Omega$) is defined by $\sqrt{GM_{\star}/R_{\rm disk}^{3}}$ and the surface density ($\Sigma$) is calculated from the final total mass and emitting radius using the formula $\sqrt{M_{\rm disk}/\pi R^{2}}$. The sound speed of the gas (c$_{s}$) is calculated using $\sqrt{k_{B}T/\mu m_{H}}$ where $k_{B}$ is the boltzmann constant, T is taken to be the same temperature as used for the mass calculations (25~K), $\mu$ is the mean molecular weight which we use 2.3 \citep[i.e.][]{Armitage19}. Using the values from Table \ref{tab:SourceResults}, we find Toomre Q parameters from 0.06 - 0.79, with V1057 Cyg corresponding to the most gravitationally unstable. All sources in this sample have a Toomre Q parameter below one, suggesting that the inner regions of all of these objects contain enough mass to trigger a sudden outburst event. Previous mass determinations of FU Ori sorces also find that the majority of these objects tend to be gravitationally unstable \citep{Cieza18, Kospal21}.

Due to our beam-size, we are unable to resolve disk structure, and our continuum observations are unresolved on scales of a typical disk. Additionally, moment~1 maps of \ce{C^{17}O} showing the intensity weighted average velocity do not show typical signatures of disk rotation.  Thus we are unable to make strong constraints on the protoplanetary disk size. Given our beam sizes of $\sim$1000 au, we are also unable to see clear Keplarian rotation from the disk in a position-velocity diagram. However, with such a relatively high mass within $<$1000~au diameter region around each source, our results are consistent with previous work in determine FU Ori-type object central masses. What is clear about these systems is that they are full of complex structure. The moment 0 maps of \ce{^{13}CO} and \ce{^{18}CO} for each source show extended emission, and for three sources seem to suggest asymmetric features that could be part of infalling structures or streamers (V1057 Cyg, V2492 Cyg, and V1735 Cyg). 
There is very likely larger scale emission that is resolved out due to our beam size, and follow up work using single dish data will be able to focus on the large scale environment of each source. The triggering of these outbursting sources may not only rely on the disk mass and dynamics, but also the surrounding environment and infalling material.

\section{Conclusion}\label{sec:conclusion}

Using the NOEMA interferometer, we observed five outbursting sources, three FU Ori objects and two EX Ori/Peculiar sources. \ce{C^{17}O} was detected in each source and was found to be optically thin, thus a mass tracer. We found masses of the centrally peaked clumps from 0.31-3.4 M$_{\odot}$ using both dust and \CsvO{} emission to determine a range of masses for the central region  of each source (along the line of sight with a radial extent of $<$1000~au around the central star).  While these masses seem to represent upper limits on the total disk mass, it falls in line with a general trend seen in FU Ori objects suggesting massive disks that may be gravitationally unstable. 


\section{Acknowledgments}

Matplotlib \citep{Matplotlib}, Astropy \citep{astropy:2013,astropy:2018}, NumPy \citep{harris2020array}

J.K.C. acknowledges support from the National Science Foundation Graduate Research Fellowship under Grant No. DGE 1256260 and the National Aeronautics and Space Administration FINESST grant, under Grant no. 80NSSC19K1534. 

E.A.B. acknowledge support from NSF Grant\#1907653 and NASA grant XRP 80NSSC20K0259





\bibliography{references}{}
\bibliographystyle{aasjournal}

\newpage

\end{document}